\documentclass[proceedings]{JHEP3}

\PrHEP{PrHEP hep2001}
\conference{International Europhysics Conference on HEP}

\usepackage{epsfig}

\def\ev{\,{\rm eV}}
\def\gev{\,{\rm GeV}}

\def\rmd{{\rm d}}

\def\b{{\rm b}}

\def\EW{{\rm EW}}

\def\L{{\rm L}}
\def\m{{\rm m}}
\def\N{{\rm N}}
\def\P{{\rm P}}

\def\etal{{\it et al.}}

\title{New results in cosmology}

\author{\speaker{Subir Sarkar}\\
        Department of Physics, 1 Keble Road, Oxford OX1 3NP, UK\\
        E-mail: \email{s.sarkar@physics.ox.ac.uk}}

\abstract{From an observational perspective cosmology is today in
excellent shape --- advances in instrumentation and data processing
have enabled us to study the universe in detail back to when the first
galaxies formed, map the fluctuations in the cosmic microwave
background which provide a measure of the overall geometry, and
reconstruct the thermal history reliably back to at least the
primordial nucleosynthesis era. However recent deep studies of the
Hubble expansion rate have suggested that the universe is
accelerating, driven by some form of `dark' (vacuum) energy. If true,
this implies a new energy scale in Nature of ${\cal O}(10^{-3})\ev$,
well below any known scale of fundamental physics. This has refocussed
attention on the notorious cosmological constant problem at the
interface of general relativity and quantum field theory. It is
possible that the resolution of this situation will require
fundamental modifications to our ideas about gravity.}

\begin{document}

\section{Introduction}

Last year, at the Osaka HEP conference, Fukugita \cite{fuk} reviewed
the emerging evidence for the possibility that the evolution of the
universe today is being driven by a cosmological constant ---
Einstein's ``greatest blunder'' \cite{ein}. In the intervening period
there has been growing acceptance of this amazing claim by the
astronomical community, and even by many particle physicists. If true,
the implied small but non-zero energy density of the present vacuum
poses a deep mystery for any fundamental theory \cite{wei}.

In the Standard Model Lagrangian, the term corresponding to the
cosmological constant is one of the two `super-renormalisable' terms
allowed by the gauge symmetries, the second one being the quadratic
divergence in the mass of fundamental scalar fields due to radiative
corrections \cite{zwi}. To tame the latter sufficiently in order to
explain the experimental success of the Standard Model has required
the introduction of a supersymmetry between bosonic and fermionic
fields which is (softly) broken at about the Fermi scale. Thus the
cutoff scale of the Standard Model, viewed as an effective field
theory, can be lowered from the Planck scale,
$M_\P\equiv\,(8\pi\,G_\N)^{-1/2}\simeq2.4\times10^{18}\gev$, down to
the Fermi scale, $M_\EW\sim\,G_{\rm F}^{-1/2}\sim300\,\gev$, albeit at
the expense of introducing over 150 new masses and other parameters in
the Lagrangian (as well as requiring delicate control of the many
non-renormalisable operators which can generate flavour-changing
neutral currents, nucleon decay etc, so as not to violate experimental
bounds). This implies a {\em minimum} contribution to the vacuum
energy density from quantum fluctuations of ${\cal O}(M_\EW^4)$,
i.e. ``halfway'' down (on a logarithmic scale) from the Planck scale
towards the energy scale of ${\cal O}(M_\EW^2/M_\P)$ corresponding to
the observationally indicated vacuum energy. Thus even the
introduction of supersymmetry cannot eradicate a discrepancy by at
least a factor of $\sim10^{60}$ between expectations and observations.

It is generally recognised that a true resolution of the cosmological
constant problem can only be achieved in a full quantum theory of
gravity. Recent developments in string theory and the possibility that
there exist new dimensions in Nature have generated many interesting
ideas concerning possible values of the cosmological constant
\cite{rev}. However it remains true that there is {\em no} generally
accepted solution to the problem of the discrepancy referred to
above. Of course the cosmological constant problem is not new but
there had always been the expectation that somehow we would understand
one day why it is exactly zero. However if it is in fact non-zero and
dynamically important today, the crisis is even more severe since it
also raises a cosmic `coincidence' problem, viz. why is present epoch
of expansion special? The resolution of these conundrums may well
require modifications to the construction of cosmological models (and
the interpretation of astronomical data) in terms of Einstein's
general relativity, although to date no such alternative which is
phenomenologically satisfactory has been presented either.

Given this sorry situation on the theoretical front, I will focus in
this talk solely on the new observational developments and in
particular provide a critical assessment of the evidence for a
cosmological constant. For lack of time I will not discuss the
continuing theoretical activity in early universe cosmology concerning
baryogenesis, inflation, particle dark matter etc. The most exciting
development in this area has been the realisation that if there are
new dimensions in Nature then they may have dramatically altered the
evolution of the early universe so the standard lore \cite{kol} may
have to be substantially rewritten. However attempts to constrain such
new physics from cosmological arguments have also brought home the
fact that we still have no detailed understanding of the early
universe before the big bang nucleosynthesis (BBN) era begining at
$\sim1$~s. In principle we know the physics underlying the evolution
back to the epoch of electroweak symmetry breaking at $\sim10^{-12}$~s
but it appears that neither this event, nor the subsequent QCD
confinement at $\sim10^{-6}$~s, would have left any detectable relics
\cite{lai}. Thus early universe theorists presently have a free hand
in speculating about the thermal history before the BBN era, in
attempting to account for the known relics, viz. the baryon asymmetry,
the abundance of dark matter and the density perturbation which seeded
the growth of large-scale structure (LSS) and left its imprint on the
cosmic microwave background (CMB). Hopefully the rapid observational
progress being made on the cosmological front, particularly in
precision studies of the CMB, will provide valuable constraints on
(and possibly clues to) the physical processes that created our
observable universe and all it contains.

\section{The observational situation}

That we live in an universe which has been hotter in the past has
recently been confirmed directly with the measurement of a higher
temperature for the CMB at high redshift. This is done by
observing fine-structure transitions between atomic levels of C~I in
quasar absorption systems (QAS) --- cold gas clouds along the line of
sight to distant quasars. Such measurements have already been possible
for some time but the inferred temperatures have had to be considered
as upper limits because alternative excitation mechanisms (i.e. other
than being immersed in a Planckian radiation bath) may have
contributed to the observed level populations. Recently an {\em
absolute} CMB temperature measurement of $T_{\rm CMB}=10\pm4$~K has
been made in a QAS at redshift $z=2.338$ \cite{sri} by constraining
such competing mechanisms through simultaneous measurements of the
rotational levels in H$_2$. (Subsequently a possible problem with this
measurement has been noted but an even better constrained measurement
of $T_{\rm CMB}=12.1^{+1.7}_{-8.2}$~K at $z=3.025$ has been reported
\cite{mol}.) As shown in Figure~\ref{T-z}, all such measurements agree
with the expectation in a Big Bang cosmology that the blackbody
temperature of the CMB increases with the redshift $\propto(1+z)$. This
is difficult to accomodate within the ``Quasi-Steady State Cosmology'' in
which the CMB arises through thermalisation of starlight \cite{bur}
--- a mechanism that was already severely constrained by the closeness
of the observed CMB spectrum to the Planck form \cite{pee}. The
prospects for constraining other alternative cosmologies by further
such measurements have been discussed in detail \cite{los}.

\DOUBLEFIGURE{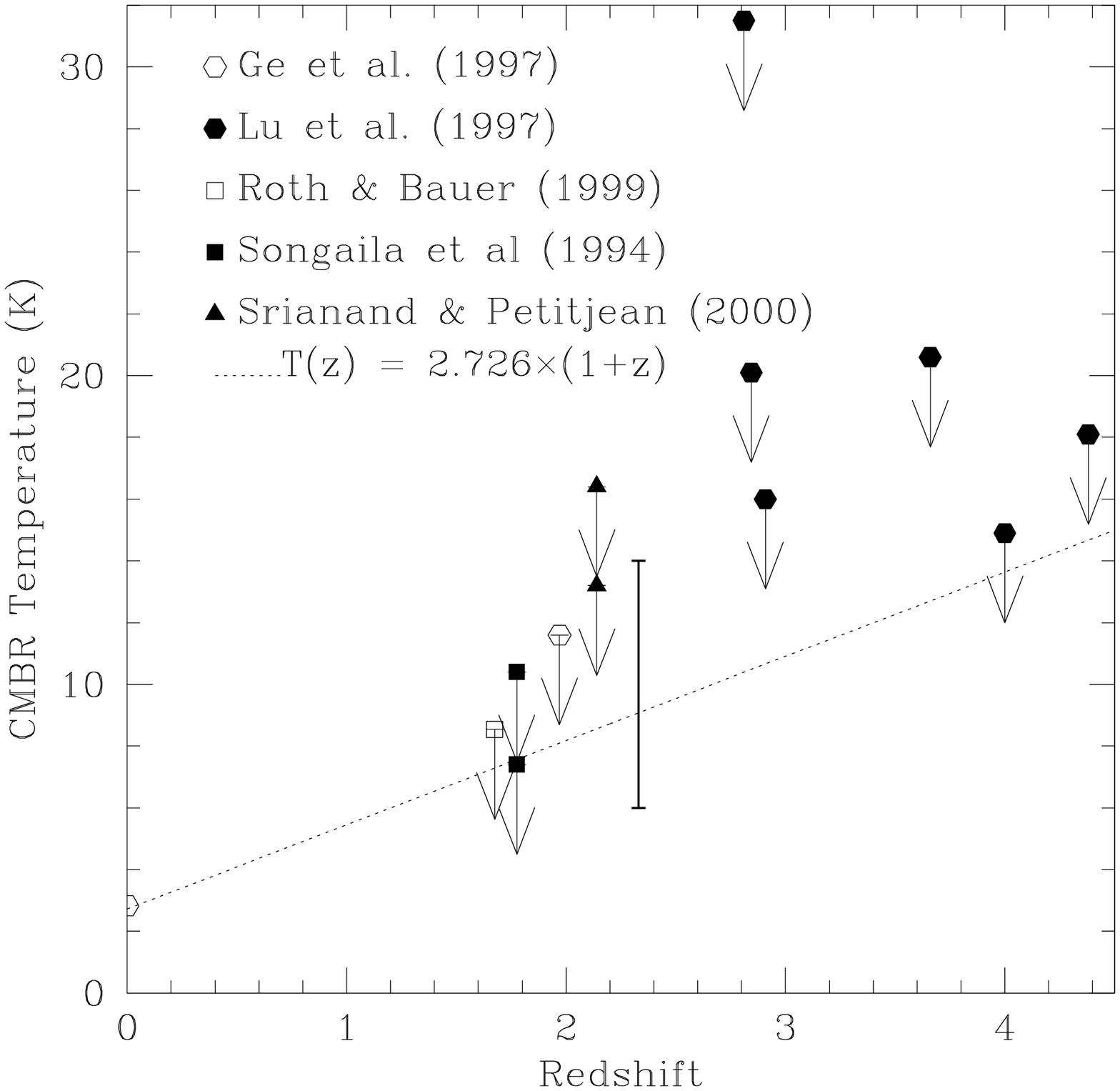,width=7cm}{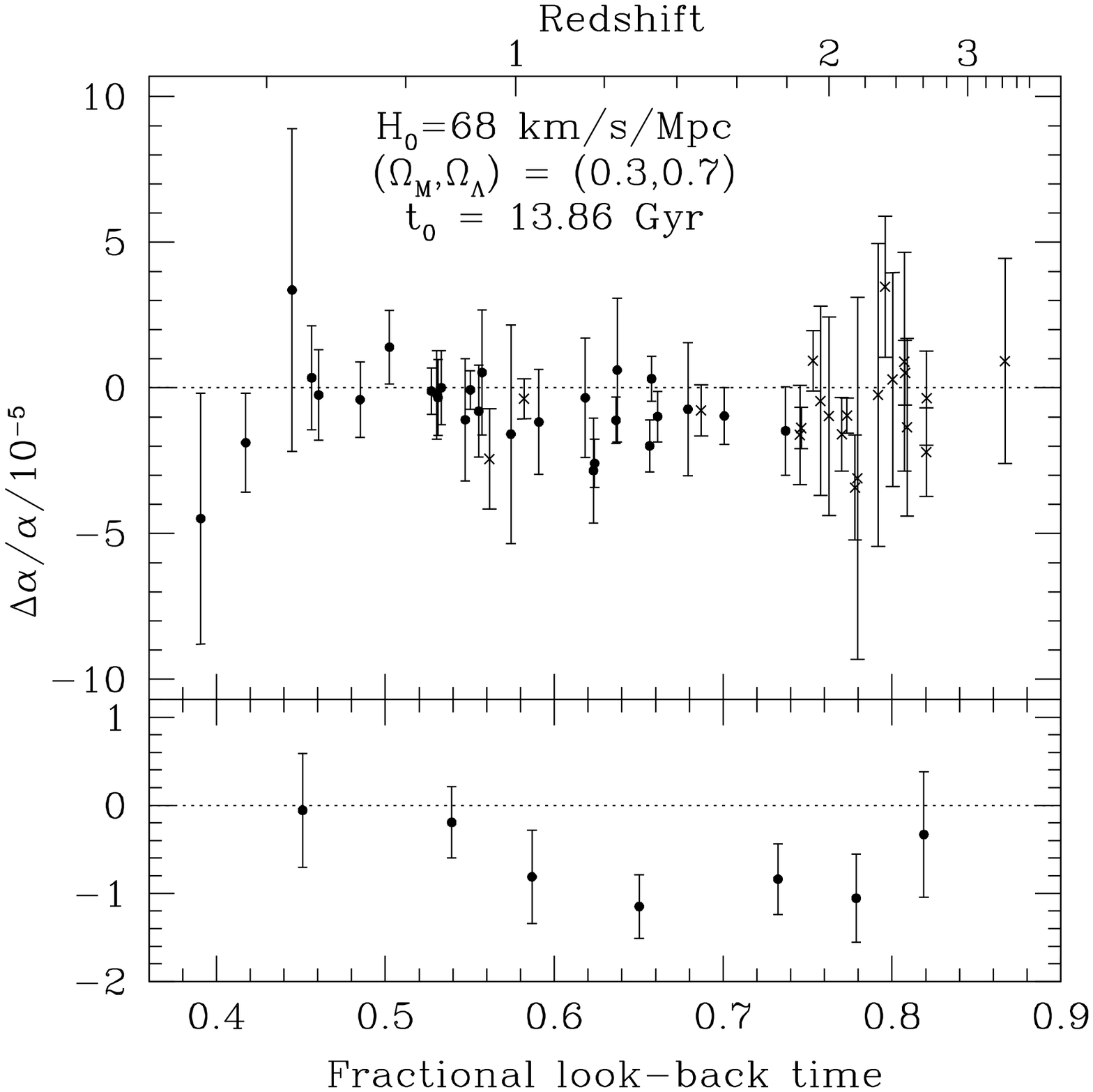,width=7cm}{Measurements
of the CMB temperature at various redshifts compared with the standard
expectation (dotted line) normalised to the COBE measurement at $z=0$
(from Ref.\protect\cite{sri}).\label{T-z}}{Inferred time variation in
the fine-structure constant from observations of QAS. The bottom panel
shows an arbitrary binning of the raw results in the upper panel
(from Ref.\protect\cite{mur}).\label{alpha-z}}

For example such measurements also imply constraints on
time-variations of fundamental constants, in particular of the
fine-structure constant $\alpha$. This is a possibility that has
attracted much theoretical interest \cite{mar,car} and very recently a
positive observational result has been claimed \cite{web,mur}. Four
independent samples were obtained (using three optical data sets and
two 21 cm and mm absorption systems) spanning upto $\sim90\%$ of the
age of the universe; each sample yields a smaller value of $\alpha$ in
the past \cite{web}. As shown in Figure~\ref{alpha-z}, the optical
sample of 49 QAS shows a $4\sigma$ deviation:
$\Delta\alpha/\alpha=(-0.72\pm0.18)\times10^{-5}$ over the redshift
range $z\sim0.5-3.5$. Moreover it is argued that correcting for any
possible systematic effects would {\em increase} the significance of
the deviation of $\Delta\alpha/\alpha$ from zero \cite{mur}. If true
this is a most exciting result since it implies the existence (in
4-dim effective field theory) of a very weakly coupled ultralight
scalar field which would mediate a `fifth force' \cite{car,dva}; such
a field is not stable under renormalisation and would require massive
fine-tuning, further exacerbating the cosmological constant problem
\cite{ban}. However studies of the 21 cm line of hydrogen yield a
conflicting bound at $z=1.6$ of
$\Delta\alpha/\alpha=(3.5\pm5.5)\times10^{-6}$ \cite{cow}. Moreover
c onsiderations of BBN place an upper limit of $\lesssim1\%$ on
$\Delta\alpha/\alpha$ at $z\sim10^{10}$ \cite{bbnalpha} and this rules
out some varying-$\alpha$ cosmologies \cite{kab}, although not others
\cite{bsm}.

\subsection{The age of the universe}

Returning to the standard cosmology, another recent observational
highlight has been the direct measurement of the age of an extremely
metal-poor (i.e. very old) star in the halo of our Galaxy through
detection of the 385.957 nm line of singly ionized $^{238}$U, as shown
in Figure~\ref{u238} \cite{cay}. (In fact $^{232}$Th ($t_{1/2}=14$~
Gyr) had already been detected in similar stars and used to infer an
age of $15.6\pm4.6$~Gyr \cite{tru} but it decays so little over the
lifetime of the Universe that $^{238}$U ($t_{1/2}=4.5$~Gyr) is in
principle a more sensitive probe.) The derived abundance,
log(U/H)=$-13.7\pm0.14\pm0.12$ corresponds to an age of
$12.5\pm3$~Gyr, consistent with the (recently revised) age of
$11.5\pm1.3$~Gyr for the oldest stars in globular clusters inferred
from stellar evolution arguments \cite{cha}. Although concerns remain
about the internal consistency of this result, e.g. the observed Th/Eu
ratio implies a much smaller age for this star \cite{tru}, direct
radioactive dating avoids many of the uncertainties that have plagued
estimates of globular cluster ages using the ``Main Sequence Turn-off
Method'', e.g. the inferred values had to be revised downward by
several Gyr following the {\sl Hipparcos} calibration of stellar
distances obtained by parallax measurements \cite{cha}. Other methods
which determine the distance and the age simultaneously (e.g. using
the ``Luminosity Function'', see Figure~\ref{globclus}) have been
developed and yield a minimum age for globular clusters of 10.5~Gyr
\cite{jim}. To obtain the age of the universe we must add
$\sim0.2-2$~Gyr, the estimated epoch of galaxy/star formation, to the
above number.

\DOUBLEFIGURE{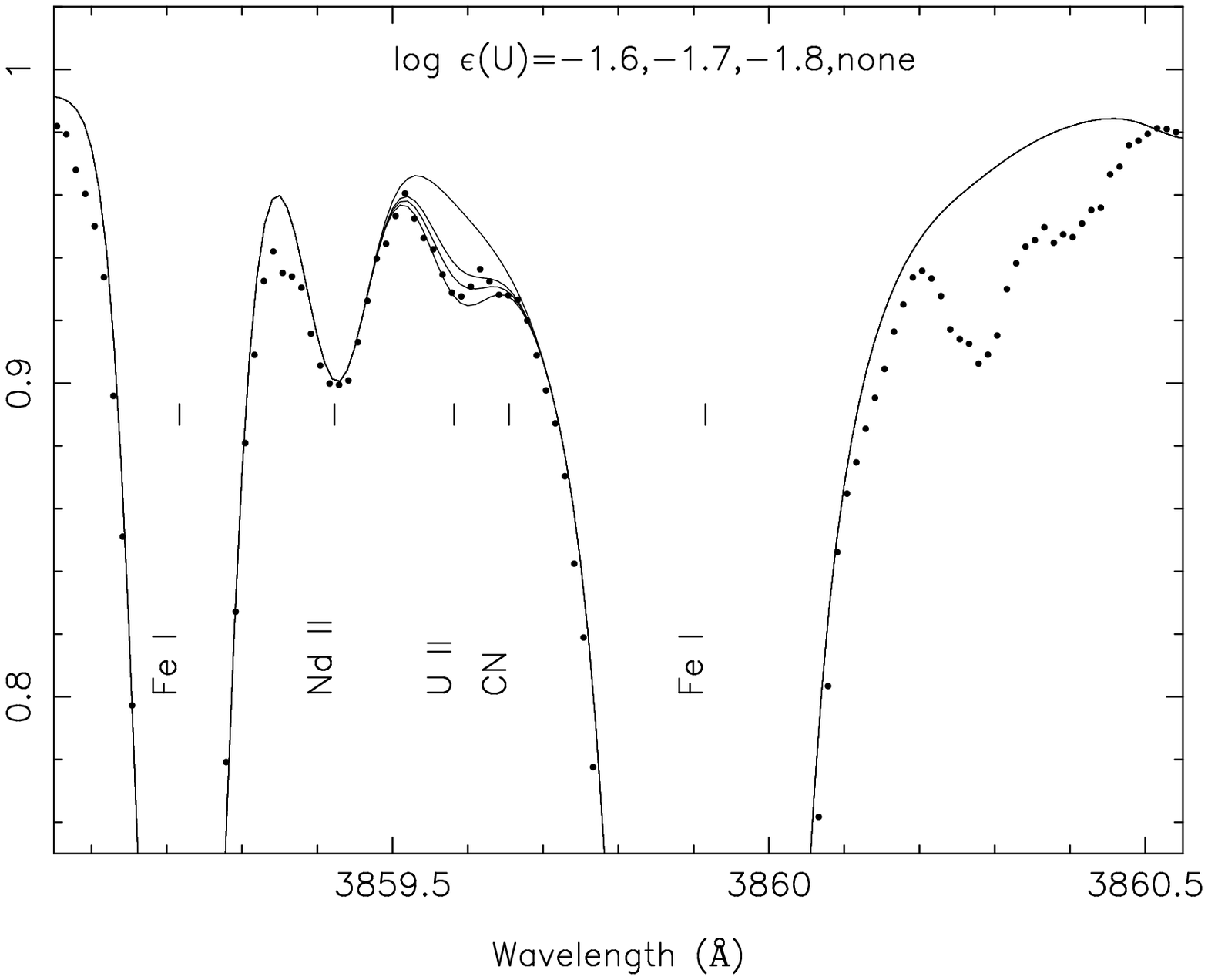,width=7cm}{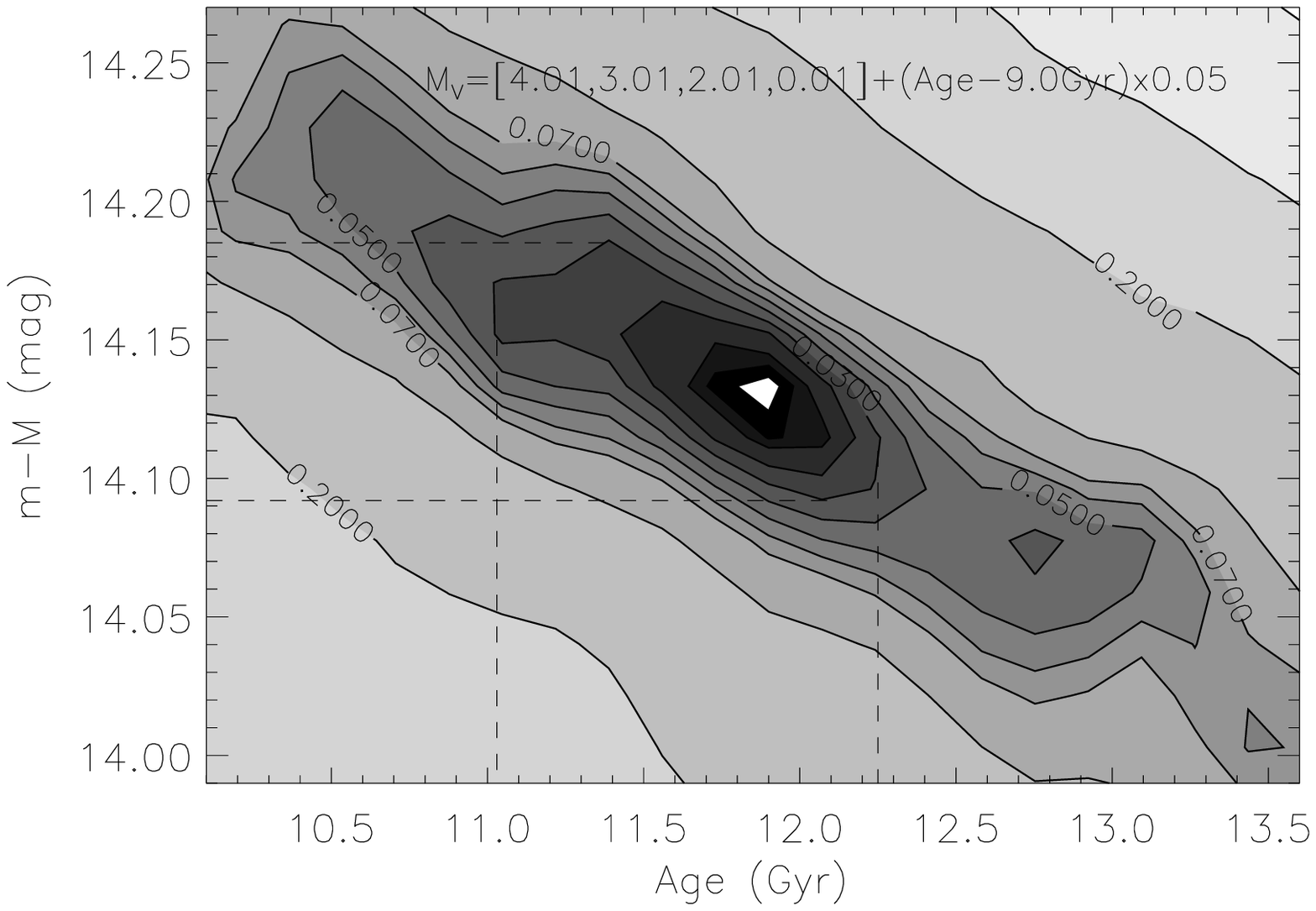,width=7.5cm}{Detection
of $^{238}$U in the old halo star CS31802-001. Synthetic spectra for
three assumed values of the abundance are compared with the data (from
Ref.\protect\cite{cay}).\label{u238}}{Simultaneous determination of
the distance modulus and age for the globular cluster M55 by the
``Luminosity Function'' method (from
Ref.\protect\cite{jim}).\label{globclus}}

\subsection{The Hubble constant}

The age thus obtained must be consistent with the expansion age of the
universe, derived from measurements of the present Hubble expansion
rate. This year the {\sl Hubble Space Telescope} Key Project has
presented its final results on $H_0$ \cite{fre}. Based on direct
measurements of the distances to 18 nearby spiral galaxies (using
Cepheid variables) and using these to calibrate five secondary
methods, they find that all data are consistent with
$H_0=72\pm3\pm7~{\rm km}~{\rm s}^{-1}~{\rm Mpc}^{-1}$, as shown in
Figure~\ref{hkp}. However significantly smaller values of $H_0$ are
obtained by `physical' methods e.g. measurements of time delays in
gravitationally lensed systems, and the Sunyaev-Zeldovich effect (SZE)
in X-ray emitting galaxy clusters, which bypass the traditional
`distance ladder' and probe far deeper distances than the objects used
by the Key Project. At present ten multiply-imaged quasars have
measured time delays or are being monitored --- the two best
constrained systems (PG1115+080 and B1608+656) yield $H_0=61\pm11~{\rm
km}~{\rm s}^{-1}~{\rm Mpc}^{-1}$ using a non-parametric modelling of
the lenses \cite{wil}, while similar low values are found (with larger
possible lens modelling uncertainties) in three other systems
\cite{koo}. Measurements of the SZE in 14 clusters also indicate a
value of $H_0\sim60~{\rm km}~{\rm s}^{-1}~{\rm Mpc}^{-1}$ with
prsently a large ($\sim30\%$) systematic uncertainty
\cite{sze}. Rowan-Robinson has argued that the Key Project data need
to be corrected for local peculiar motions using a more sophisticated
flow model than was actually used, and also for metallicity effects on
the Cepheid calibration --- this lowers the value of $H_0$ inferred
from the same dataset to $63\pm6~{\rm km}~{\rm s}^{-1}~{\rm Mpc}^{-1}$
\cite{row}.

\DOUBLEFIGURE{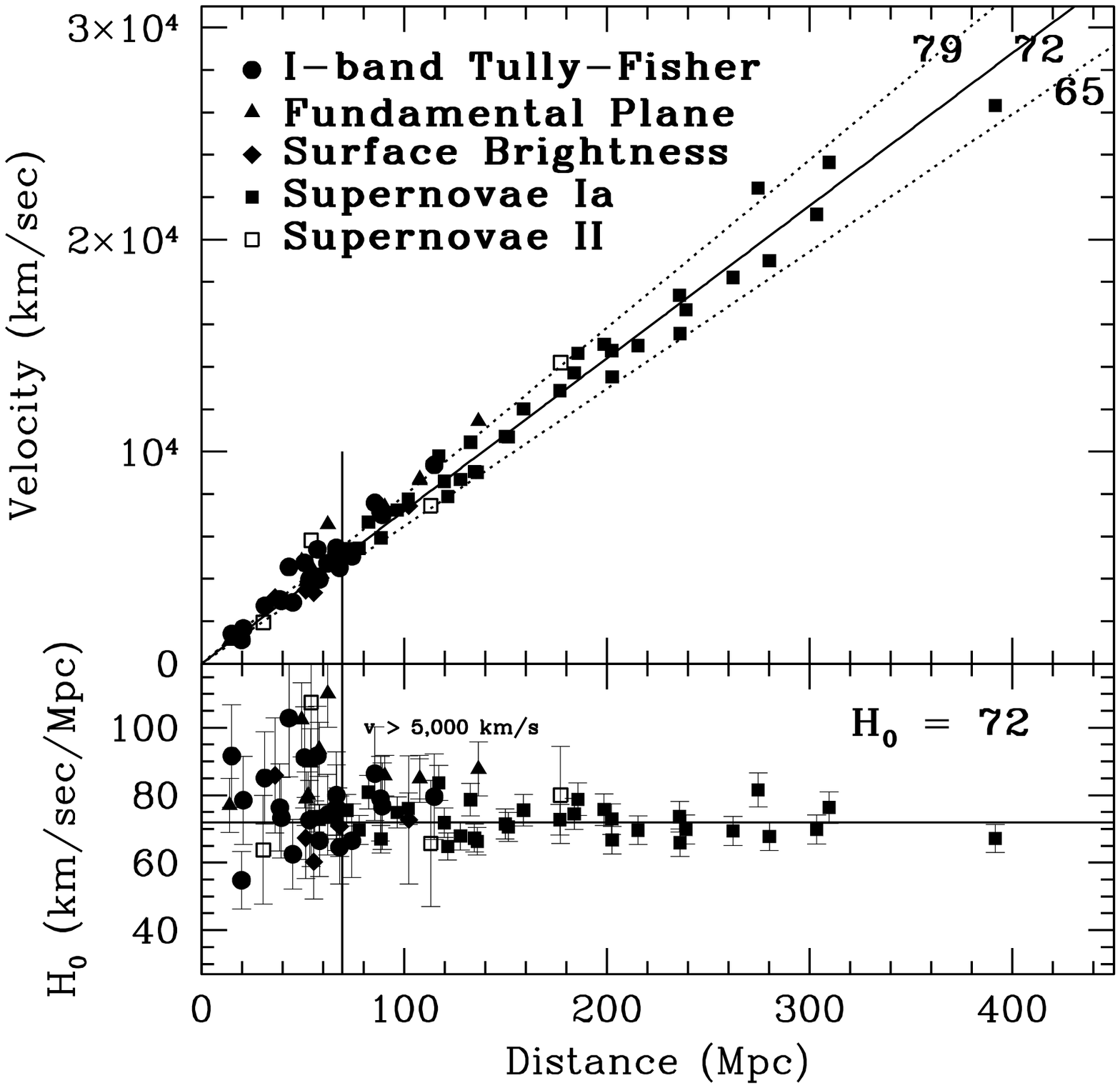,width=7.5cm}{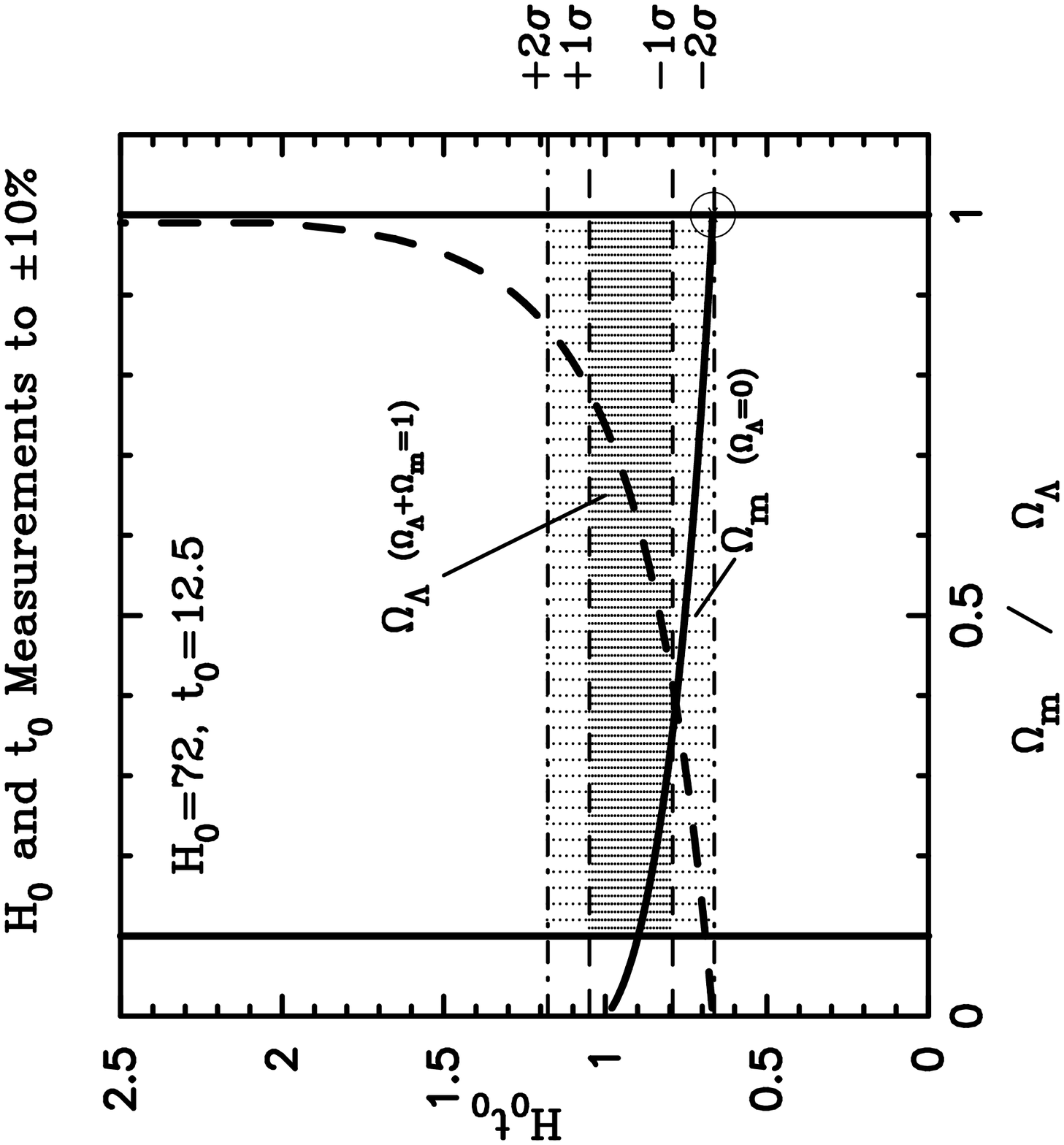,width=7cm,angle=-90}{Hubble
diagram for Cepheid-calibrated secondary distance indicators; the
bottom panel illustrates the decrease in fluctuations (due to peculiar
velocities) with increasing distance (from
Ref.\protect\cite{fre}).\label{hkp}}{The time-scale problem for an
E-deS universe (marked with a circle), assuming $\pm10\%$
uncertainties for the indicated values of the present expansion rate
and age (from Ref.\protect\cite{fre}).\label{h0t0}}

It is often stated that an Einstein-DeSitter (E-deS) universe with
$\Omega_\m=1,\Omega_\Lambda=0$ is too short-lived to be compatible
with measurements of $H_0$ and $t_0$.\footnote{Here $\Omega_\m$ and
$\Omega_\Lambda$ are the energy densities of matter and the vacuum in
units of the critical density $\rho_{\rm
c}\equiv\,3H_0^2/8{\pi}G_\N\simeq(3\times10^{-3}\ev)^4$, i.e.
$\Omega_m+\Omega_\Lambda=1-\kappa$ where the curvature term $\kappa$
is 0 for a flat universe.} As seen in Figure~\ref{h0t0}, adopting the Key
Project measurement of the present expansion rate and a reasonable age
for the universe does rule out this model at the $2\sigma$ level and
favours an open universe with $\Omega_\m\lesssim0.3$ or a flat
universe with a cosmological constant. However in view of the
continuing debate concerning the value of $H_0$, this conclusion
cannot as yet be considered definitive.

\subsection{The deceleration parameter}

The most exciting observational developments have undoubtedly been in
measurements of the deceleration parameter
$q\equiv{\rmd}H^{-1}/{\rmd}t-1$, which equals 0.5 for the E-deS model
where $H\propto\,t^{2/3}$, and -1 for a DeSitter (deS) model with
$\Omega_\m=0,\Omega_\Lambda=1$ which has $H$ constant. This has been
done through impressive deep studies of the Hubble diagram of Type~Ia
supernovae (SNe~Ia) carried out by the {\sl Supernova Cosmology
Project} (SCP) \cite{scp} and the {\sl High-z SN Search Team} (HZT)
\cite{hzt}. Their basic observation is that distant supernovae (upto
$z\sim1$ corresponding to looking back $\sim10$~Gyr) are upto
$\Delta{m}\sim0.5$ mag (corresponding to
$10^{\Delta{m}/2.5}-1\simeq60\%$) fainter than would be expected for a
decelerating universe such as the E-deS model. This has been
interpreted as implying that the expansion rate has been {\em speeding
up} since then, thus the observed SNe~Ia are actually further away
than expected.\footnote{The measured apparent magnitude $m$ of a
source of known absolute magnitude $M$ yields the `luminosity
distance': $m-M=5\log(\frac{d_\L}{\rm Mpc})+25$, where
$d_\L=(1+z)\int_0^z\frac{{\rmd}z'}{H(z')}$ is sensitive to the
expansion history \cite{car2}.} The obvious possibility that the
SNe~Ia appear fainter because of absorption by intervening dust can be
constrained since this would also lead to reddening (unless the dust
has unusual properties \cite{agu}). It is more difficult to rule out
the possibility of evolution, i.e. that the distant SNe~Ia are
intrinsically fainter. Many careful analyses have been made of these
possibilities by the observing teams themselves as well as by
others. A detailed review of the data and their implications for
cosmology has been given by Leibundgut \cite{lei}; the summary of the
present situation below draws largely on this work.

Briefly, SNe~Ia are observationally known to be a rather homogeneous
class of objects, with intrinsic peak luminosity variations
$\lesssim20\%$, hence particularly well suited for cosmological tests
which require a `standard candle' \cite{bra}. They are characterised
by the absence of hydrogen in their spectra \cite{fil} and are
believed to result from the thermonuclear explosion of a white dwarf,
although there is as yet no ``standard model'' for the progenitor(s)
\cite{hil}. However it is known (using nearby objects with
independently known distances) that the time evolution of SNe~Ia is
tightly correlated with their peak luminosities such that the
intrinsically brighter ones fade faster --- this can be used to make
corrections to reduce the scatter in the Hubble diagram using various
empirical methods, viz. the $\Delta\,m_{15}$ template \cite{phi} or
the Multi-colour Light Curve Shape (MLCS) used by the HZT \cite{hzt};
the SCP uses a simple `stretch factor' to normalise the observe
apparent peak magnitudes \cite{scp}. {\em It should be emphasised that
such corrections are essential to reduce the scatter in the data
sufficiently so as to allow significant cosmological deductions}. It
is therefore a matter for concern that, as seen in Figure~\ref{comp},
the different methods make magnitude corrections that do not compare
well with each other \cite{dre,lei} --- as they should if indeed some
intrinsic physical property of SNe~Ia was responsible for the observed
correlations. Moreover as shown in Figure~\ref{sncol}, distant SNe~Ia
appear to be bluer in colur than nearby ones; this suggests that the
derived reddening may have been underestimated \cite{lei}. A recent
reanalysis of the data also finds that both teams have underestimated
the effects of host galaxy extinction and that the peak
luminosity-decay time correlation is much weakened if SNe~Ia not
observed before maximum light are excluded; this further weakens the
significance of the claimed acceleration \cite{row2}.

\DOUBLEFIGURE{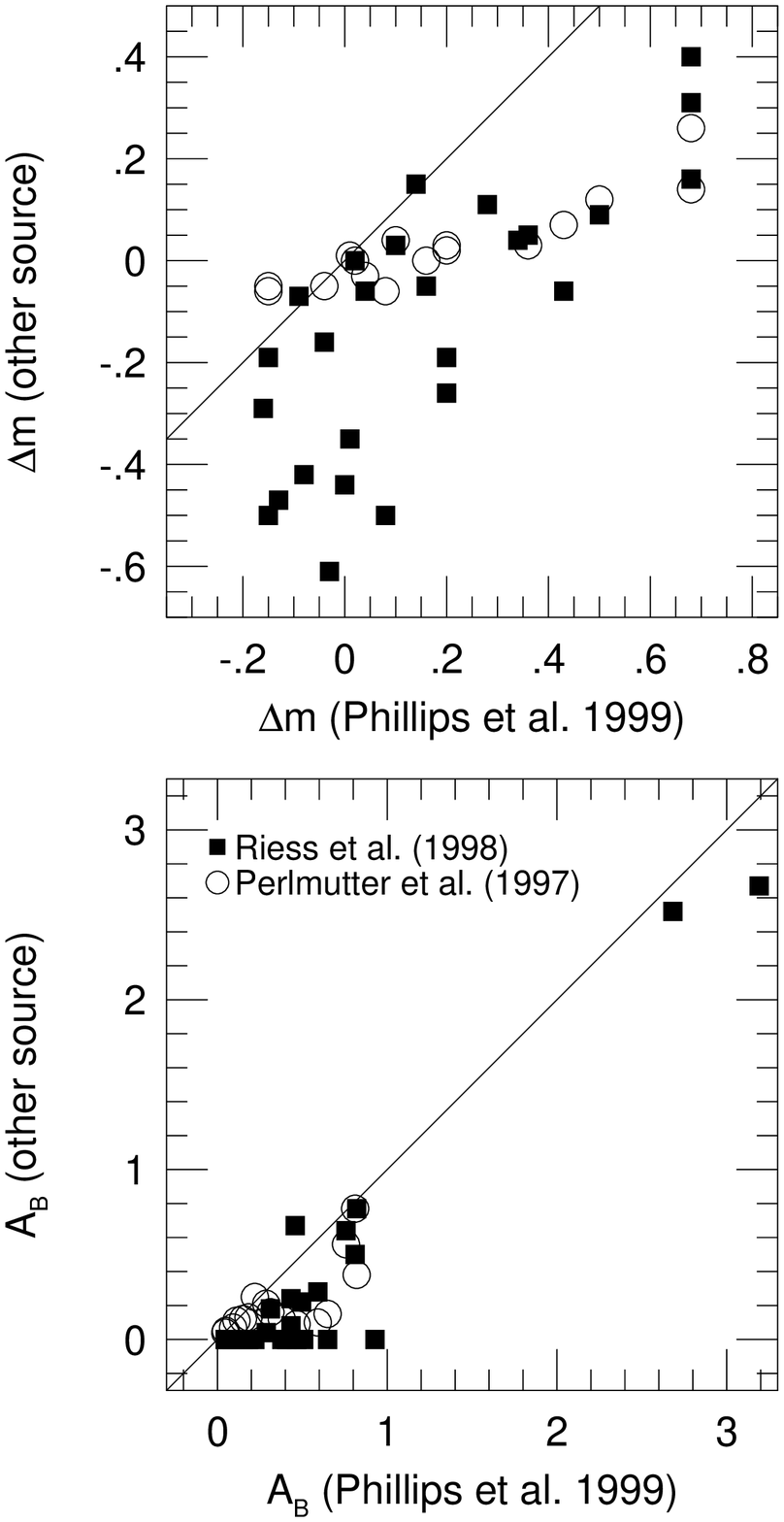,height=7cm}{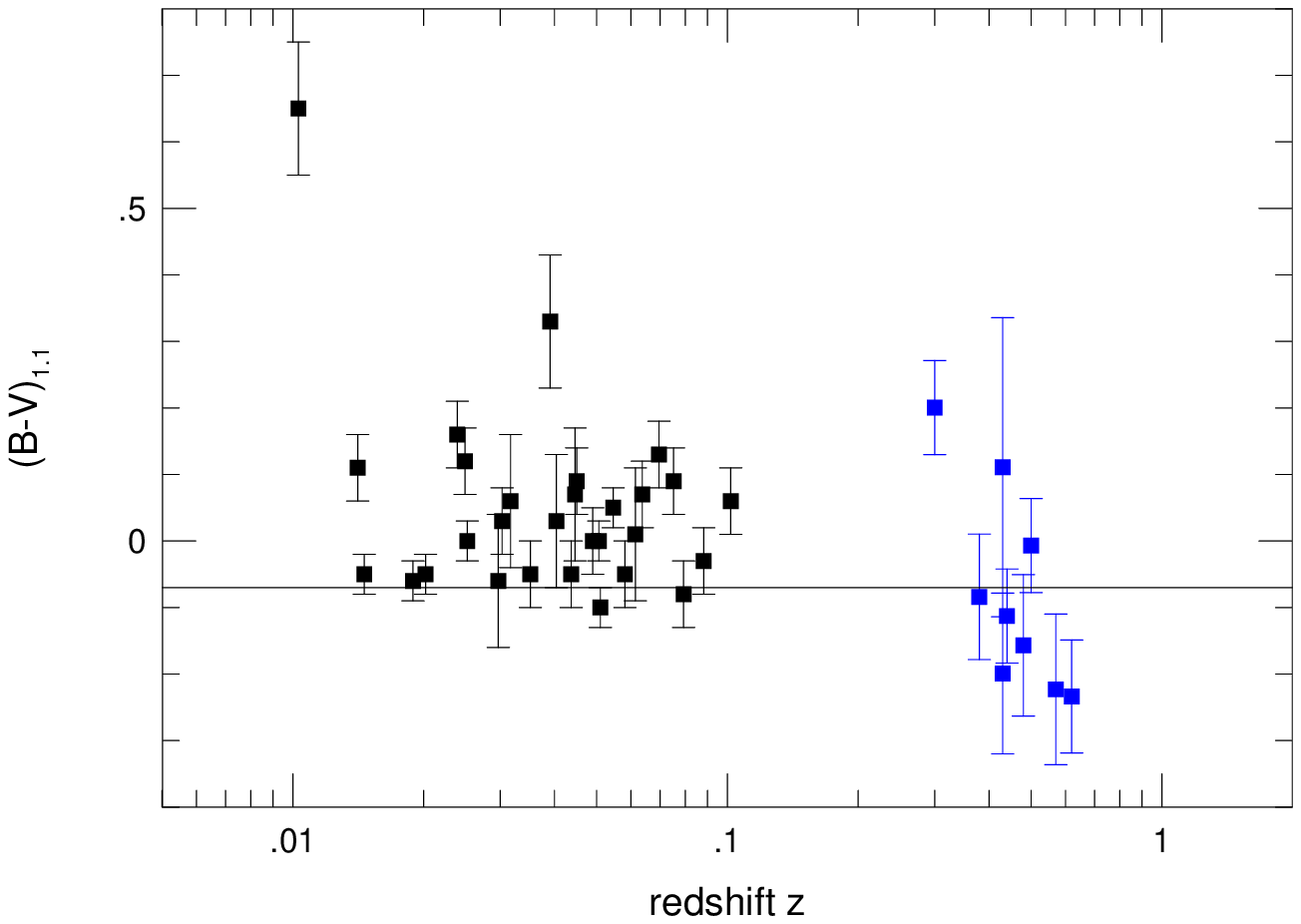,width=7cm}{Comparison of
the magnitude corrections (top panel) and of the inferred absorption
(bottom panel) for the nearby SNe~Ia sample, for the three different
methods used to analyse the data (from
Ref.\protect\cite{lei2}).\label{comp}}{The observed colours of low
redshift \protect\cite{phi} vs high redshift \protect\cite{hzt} SNe~Ia
(from Ref.\protect\cite{lei}).\label{sncol}}

Figure~\ref{snhub} shows the magnitude-redshift diagram of SNe~Ia
obtained by the two teams. Averaging over the distant supernovae, it
is found that they are $0.20\pm0.06$ magnitudes fainter at $z\sim0.5$
than in an (empty) universe expanding at constant rate when analysed
using the $\Delta\,m_{15}$ correction method; this decreases to
$0.14\pm0.06$ using the MLCS method as done by HZT \cite{hzt}, and to
only $0.06\pm0.04$ for the SCP sample \cite{scp}. Thus the claim that
cosmic acceleration has been observed is no more significant than the
recent indications for the Higgs at LEP! The best fit `$\Lambda$CDM
model' --- a low matter density, cosmological constant dominated flat
universe with $\Omega_\m\simeq0.3,\Omega_\Lambda\simeq0.7$ --- is
however favoured by measurements of the age and present expansion rate
as discussed earlier, and also by independent measurements of the
matter density and spatial curvature to be discussed, so has become
the new ``standard cosmological model'' for the astronomical
community.

\EPSFIGURE{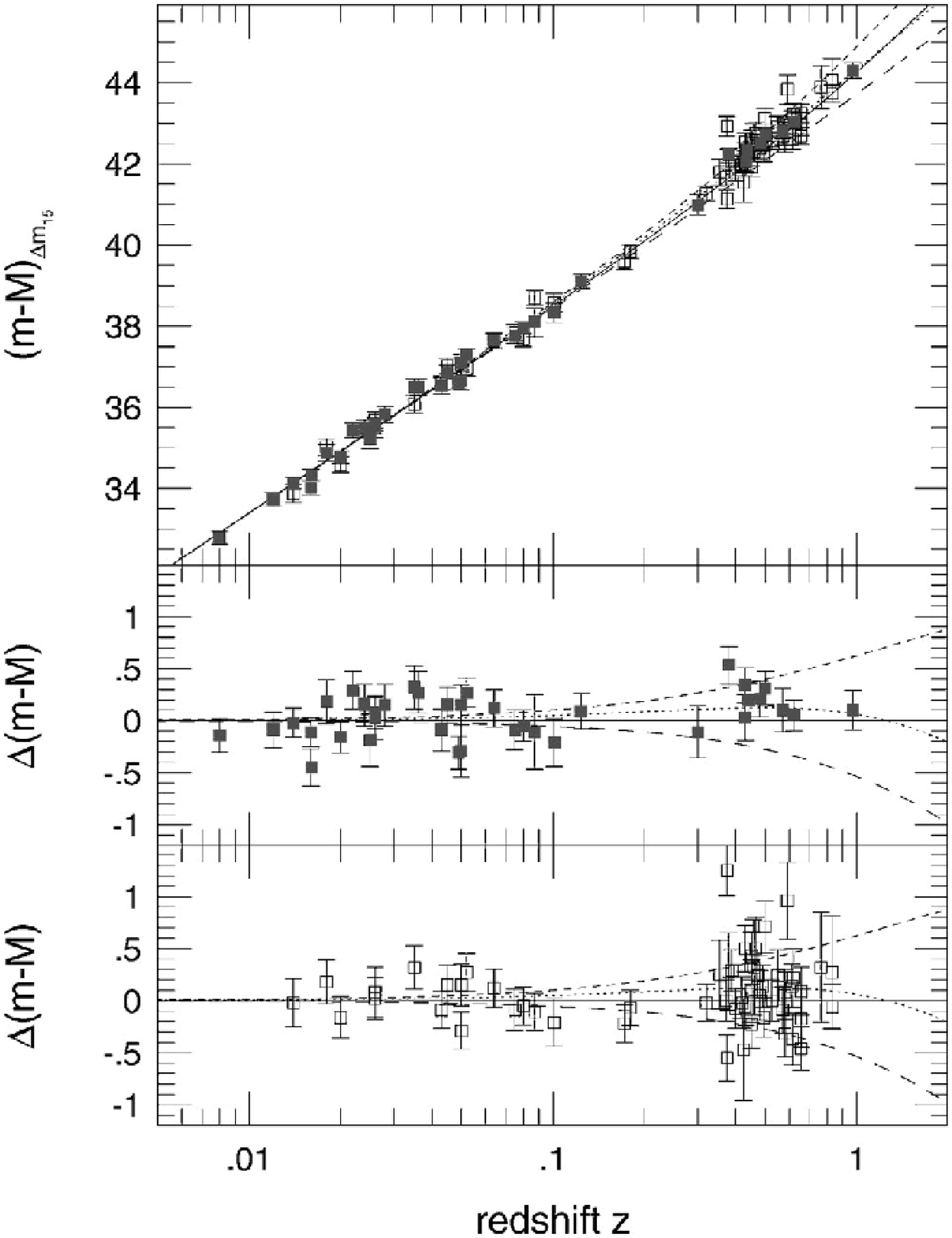,width=10cm} {The Hubble diagram of SNe~Ia (top
panel) compared with the expectation for an empty universe (full
line), an E-deS universe $\Omega_\m=1,~\Omega_\Lambda=0$ (long-dashed
line), a DeS universe $\Omega_\m=0,~\Omega_\Lambda=1$ (dashed line),
and a flat $\Omega_\m=0.3,~\Omega_\Lambda=7$ universe (dotted line),
with all data normalised to the $\Delta\,m_{15}$ method
\protect\cite{phi}. The lowest panel shows separately the Supernova
Cosmology Project data \protect\cite{scp}, and the middle panel, the
High-z SN Search Team data \protect\cite{hzt} (from
Ref.\protect\cite{lei}).\label{snhub}}

More recently the serendipitous discovery of a very distant supernova
(SN1997ff) in the Hubble Deep Field at a redshift $z\sim1.7$ is
claimed to have largely eliminated alternative explanations such as
absorption by ``grey'' dust or luminosity evolution in favour of a
cosmological constant \cite{rie}. This is because for $z\gtrsim1$ the
cosmological constant becomes unimportant relative to the increasing
matter density ($\propto(1+z)^3$) so the expansion should be seen to
be slowing down at such epochs; the transition from acceleration to
deceleration occurs at $z=(2\Omega_\Lambda/\Omega_m)^{1/3}-1$. It is
argued \cite{rie} that SN1997ff is indeed a SNe~Ia because the host
galaxy is an elliptical rather than a spiral and this identification
is consistent with the observed colours and time evolution. As seen
in Figure~\ref{sn97ff}, the inferred luminosity then does indicate
deceleration in that SN1997ff is {\em brighter} than would be expected
for an uniform rate of expansion; this argues against absorption by
dust or simple forms of luminosity evolution as being responsible for
the curvature in the Hubble diagram at lower redshifts. However it has
been noted that SN1997ff may have been significantly brightened by
gravitational lensing due to two foreground elliptical galaxies which
are close to the line of sight \cite{sn97fflens} so this is not quite
the clean `smoking gun' signature hoped for. However it does
illustrate the effectiveness of such deep observations. The proposed
{\sl Supernova Acceleration Probe} (SNAP) space mission \cite{snap}
will observe $\sim2000$ SNe~Ia out to a redshift of $z\simeq1.2$ and
the increase in statistics and improved control of systematics should
be able to improve the accuracy of $\Omega_\m$ and $\Omega_\Lambda$
determinations to a few per cent! Only through such future
measurements can the claim of recent cosmic acceleration, driven by
vacuum energy, be definitively tested.

\EPSFIGURE{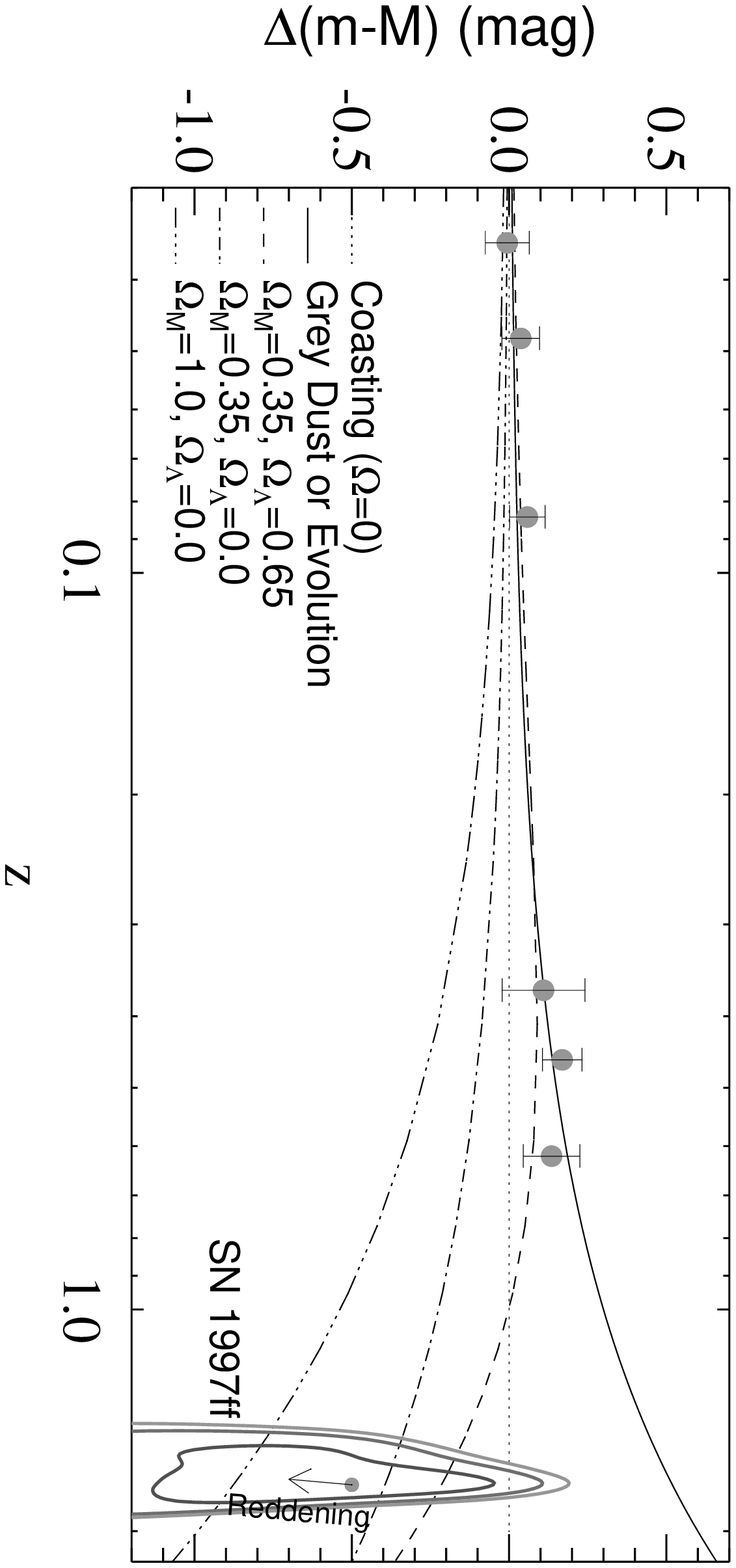,width=5cm,angle=90}{Hubble diagram of SNe~Ia
normalised to an empty uniformly expanding universe with redshift binned data
from the SCP and HZT for redshifts $\lesssim1$ and the data for
SN1997ff at $z\sim1.7$ (from
Ref.\protect\cite{rie}.\label{sn97ff}}

\section{The spatial curvature and the matter density}

Even though the direct evidence for a non-zero $\Omega_\Lambda$ is
rather weak as argued above, many astronomers have nevertheless
accepted the findings. This is really because of two {\em independent}
observations which also suggest, that there is a substantial
cosmological constant. The first, as discussed by Ganga at this
conference \cite{gan}, is that recent measurements of degree-scale
angular fluctuations in the CMB provide a measurement of the sound
horizon (a `standard rod') at last scattering \cite{cmb} and thereby
indicate that the curvature term $\kappa\simeq0$ i.e. the universe is
spatially flat \cite{wan}. The second is that, as recognised for some
time, several types of observations suggest that the amount of matter
which participates in gravitational clustering is significantly less
than the critical density, $\Omega_\m\sim0.3$ \cite{peeb,dav}. By the
`cosmic sum rule' this then requires that there be some form of `dark
energy', unclustered on the largest spatial scales probed in the
measurements of $\Omega_\m$, with an energy density of
$1-\Omega_\m\sim0.7$. This is indeed consistent with the value of
$\Omega_\Lambda\sim0.7$ suggested by the SNe~Ia data leading to the
widespread identification of the dark energy with vacuum
energy.\footnote{In seeking to understand why the vaccum energy has
begun to dominate at the present epoch many theorists have pursued the
idea that it may be the energy of an evolving scalar field ---
`quintessence' \cite{qui,qui2}. In order to be affected by the Hubble
expansion such a field must have a mass of ${\cal
O}(H_0)\sim10^{-33}\ev$, much smaller than the height of its
potential, $\sim3\times10^{-3}\ev$, thus raising formidable
difficulties in relating it to fundamental theory. However if the dark
energy is parameterised in terms of its equation of state
$w\equiv{p}/\rho$, the SNe~Ia observations are in fact best fitted by
$w\simeq-1$ \cite{per,sai} i.e. a cosmological constant!} However it
should be emphasised that this is rather indirect evidence for a
cosmological constant and there may be other explanations for the
apparent shorfall between the matter density and the critical density
required by the CMB. In view of the importance of the conclusion, each
line of evidence ought to be critically judged on its own merits.

For example the position of the first acoustic peak in the angular
power spectrum of the CMB implies that the spatial curvature is close
to zero only if the primordial scalar density perturbation, presumably
generated during inflation, is assumed to be adiabatic and
gaussian. This is indeed what is expected in simple inflationary
models \cite{inf} but given that we have no ``standard model'' of
inflation it is legitimate to consider more general perturbations,
e.g. including isocurvature modes. The most general cosmological
perturbation in an universe filled with photons, baryons, neutrinos,
and cold dark matter (CDM) has in fact several such modes including
two possible ones in the neutrino component \cite{buc}; moreover the
adiabatic and isocurvature modes are in general correlated. If such
modes exist then the inference from the position of the first peak
that the universe is flat is severely compromised \cite{buc2}. Even if
we assume that the universe is flat, the extracted values of other
cosmological parameters can be significantly altered if there are
isocurvature modes present \cite{trd}. For example, as shown in
Figures~\ref{iso1} and \ref{iso2}, the agreement of the baryon density
inferred from the CMB data with the value indicated by BBN
\cite{bbn,fie} holds only if the perturbation is adiabatic. Of course
it is not clear what physical mechanism can excite e.g. neutrino
isocurvature perturbations {\em after} neutrinos decouple at
$\sim1$~MeV, nevertheless this illustrates that ``precision''
determinations of cosmological parameters from the CMB cannot yet be
considered fully robust. To distinguish experimentally between
isocurvature and adiabatic perturbations requires measurements of the
CMB polarisation which will be possible with the forthcoming PLANCK
surveyor \cite{buc2,enq}.

\DOUBLEFIGURE{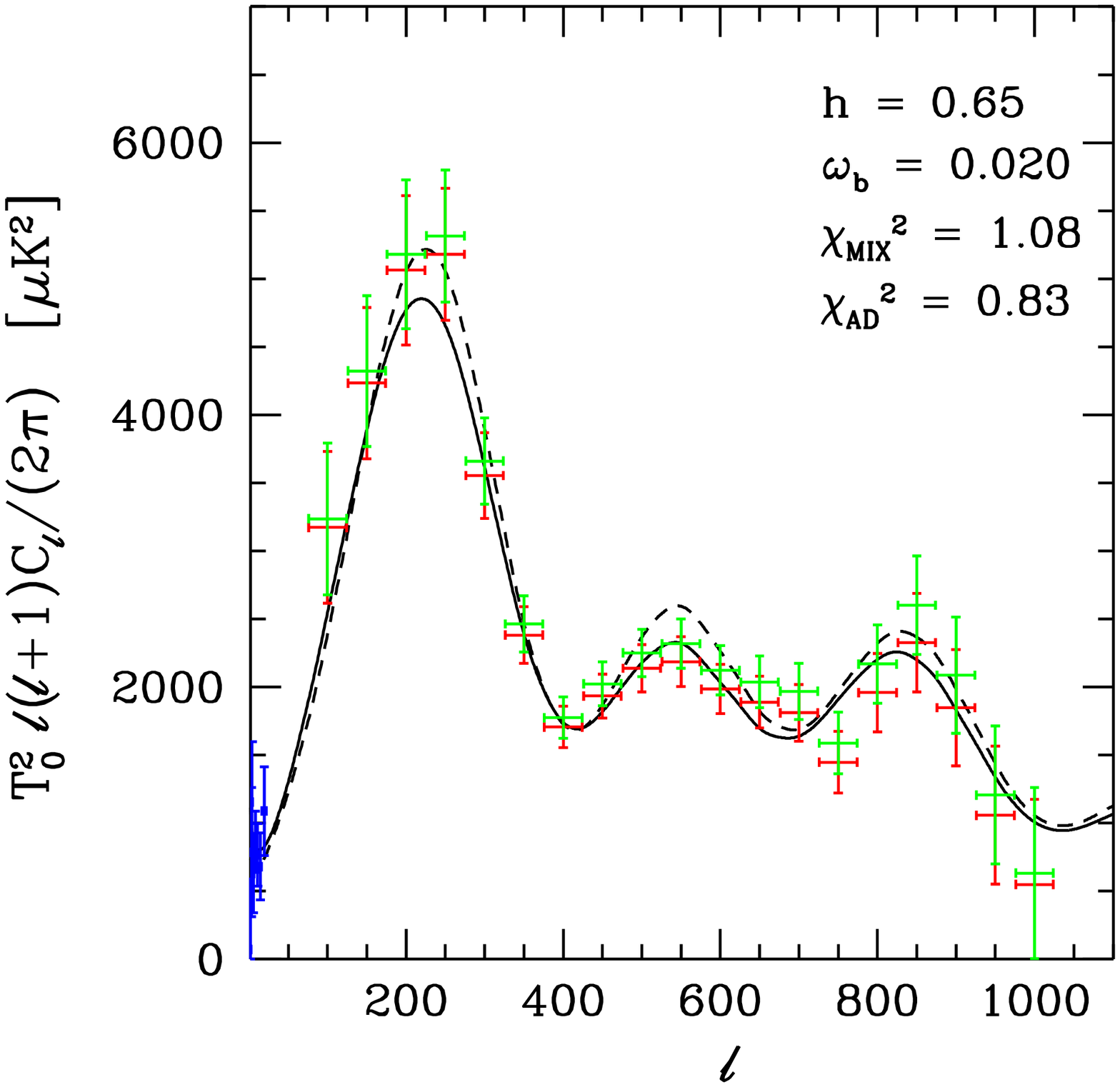,width=7cm}{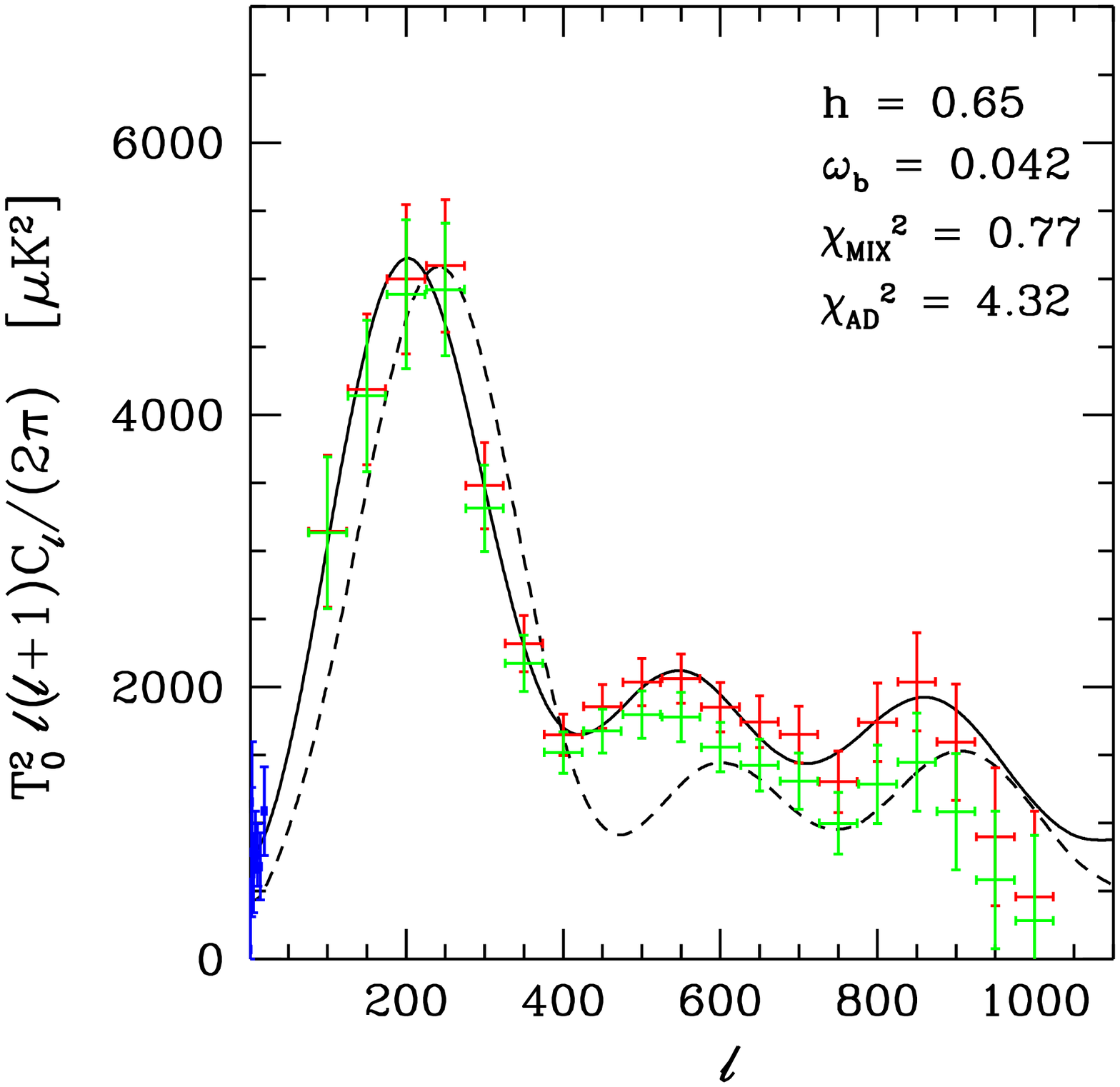,width=7cm}{Fits to the CMB
data assuming purely adiabatic perturbations (dashed line) and mixed
perturbations with $12\%$ isocurvature content (solid line), adopting
the baryon density $\omega_\b=\Omega_{\b}h^2=0.02$ indicated by
BBN.\label{iso1}}{The same but now for an assumed baryon density twice
the BBN value; the adiabatic case is now a poor fit but the mixed case
(with $69\%$ isocurvature component) is quite acceptable (from
Ref.\protect\cite{trd}).\label{iso2}}

Another important property of the primordial scalar density
perturbation about which we have no direct knowledge is its spectral
shape. It has been common practice to deduce cosmological parameters
from analyses of LSS and CMB data, {\em assuming} that the primordial
perturbation has a scale-invariant `Harrison-Zeldovich' form with
power-law index $n=1$. Indeed many of the arguments for a low matter
density universe (e.g. the `shape factor' of the LSS power spectrum)
are crucially dependent on this assumption; even allowing e.g. a small
`tilt' $n\simeq0.9$ for the spectrum allows a high matter density
universe to be consistent with the data \cite{tilt}. Moreover such a
departure from scale-invariance is indeed expected in models of `new
inflation' \cite{inf} and explicit models based on supergravity which
do yield the required tilt have been presented \cite{ros}. It is
interesting to note that consistency of the baryon density inferred
from the recent CMB data with its BBN value \cite{bbn} now does
require such a spectral tilt \cite{gan,wan}.

In fact the primordial perturbation may not even be scale-free; indeed
it is sensitive to the dynamics of inflation (and the behaviour of
fields other than the inflaton itself) and is likely to have features
such as `steps' and `bumps' \cite{fea}. Of course there is no unique
prediction in this regard so it is quite appropriate that analyses of
CMB and LSS data should still assume a scale-free spectrum (but now
allowing a `tilt' for consistency with the data). However the inferred
cosmological parameters can change substantially if we allow for a
more complex primordial spectrum, e.g. a `step' in the spectrum
permits a high matter density universe to be consistent with the LSS
and CMB power spectra \cite{bar}.\footnote{The CMB data used (from the
Boomerang experiment) has subsequently been revised but our conclusion
still holds (e.g. using the latest combined data set \cite{wan}),
although an $\Omega_\m=1$ universe no longer gives a good fit as
before. In view of the uncertainties in the measurements, particularly
at high multipoles, it might be advisable to await data from the
ongoing MAP mission \cite{map} before coming to a firm conclusion.}
The only {\em independent} way to constrain such possibilities is
through more detailed analyses of structure formation, to test the
important {\em assumption} of a scale-free (and gaussian) primordial
perturbation.

There has indeed been impressive recent progress in mapping the
present day distribution of galaxies \cite{2df,sdss} and in probing
the early `linear' epoch of structure formation at higher redshifts
through studies of the ``Lyman-$\alpha$ forest'' in QAS \cite{lya}. It
has also become possible to probe the dark matter distribution
directly through studies of the `shear' induced by gravitational
lensing \cite{mel}.  The {\em Two-degree Field} (2dF) galaxy redshift
survey \cite{2df} has detected the expected distortion in
redshift-space due to peculiar (i.e. non-Hubble) velocities induced by
the gravitational collapse of structure \cite{2df}. If galaxies do
trace the dark matter distribution, as is indicated by studies of
high-order correlation functions in the same data, then this does
provide direct evidence for a low density universe with
$\Omega_m=0.27\pm0.06$ on the largest scales (upto several hundred
Mpc) probed so far. This is also consistent with the measured power
spectrum. By doing a joint fit to the 2DF and CMB data ({\em assuming}
a scale-free primordial spectrum of adiabatic fluctuations), evidence
is found for a cosmological constant with $0.65<\Omega_\Lambda<0.85$,
\cite{efs}. With improving datasets it should soon be possible to test
the robustness of such results with even fewer assumed `priors', in
particular the nature of the primordial density perturbation, and
address other concerns about structure formation \cite{peeb}.

\section{Conclusions}

Thus for the moment there is a `cosmic concordance' model with
$\Omega_\m\sim0.3, \Omega_\Lambda\sim0.7$ which is consistent with all
astronomical data but has {\em no} explanation in terms of fundamental
physics. One might hope to eventually find explanations for the dark
matter (and baryonic) content of the universe in the context of
physics beyond the Standard Model but there appears to be little
prospect of doing so for the apparently dominant component of the
universe --- the cosmological constant. Cosmology has in the past been
a data-starved science so it has been appropriate to consider only the
simplest possible cosmological models in the framework of general
relativity. However now that we are faced with this serious
confrontation between particle physics and cosmology, it is perhaps
time to even consider possible alternatives to general relativity.

This is of course not trivial --- general relativity has been
extensively tested on (relatively small) astronomical scales
\cite{dam} and the standard cosmology based on it certainly gives a
successful account of observations back to the BBN era
\cite{fie}. However it is not unlikely that the ferment of current
theoretical ideas concerning `brane-world' might suggest small
modifications to our notions about gravity, even sacred texts like the
Equivalence Principle, which turn out to be significant in the
cosmological context \cite{dam}. Of course astronomers are entitled
to, and will continue to, analyse their data in terms of
well-established physics. But it is important for it to be recognised
that the cosmological constant is not just another parameter among
many specifying a cosmological model.

Landu famously said {\em ``Cosmologists are often wrong, but never in
doubt''}. The situation today is perhaps better captured by Pauli's
enigmatic remark --- the present interpretation of the data may be
{\it ``\ldots not even wrong''}. However we are certainly not without
doubt! Moreover the crisis posed by the recent astronomical
observations is not one that confronts cosmology alone; it is the
spectre that haunts any attempt to unite two of the most successful
creations in physics --- quantum field theory and general
relativity. The future of cosmology is inextricably intertwined with
that of fundamental physics and promises to be most exciting indeed.

\acknowledgments

I am grateful to Bruno Leibundgut and Michael Rowan-Robinson for very
helpful correspondance concerning the measurement of cosmological
parameters. I would also like to thank the organisers of this
excellent conference for the invitation to present this review.

\end{document}